# Enterprise Social Networks as Digital Infrastructures - Understanding the Utilitarian Value of Social Media at the Workplace

Christian Meske, Konstantin Wilms & Stefan Stieglitz

# Enterprise Social Networks as Digital Infrastructures - Understanding the Utilitarian Value of Social Media at the Workplace

Christian Meske [a], Konstantin Wilms[b], and Stefan Stieglitz[b]

[a]Department of Information Systems, Freie Universität Berlin and Einstein Center Digital Future, Berlin, Germany; [b]Department of Computer Science and Applied Cognitive Science, University of Duisburg-Essen, Duisburg, Germany

**ABSTRACT**
In this study, we first show that while both the perceived usefulness and perceived enjoyment of enterprise social networks impact employees' intentions for continuous participation, the utilitarian value significantly outpaces its hedonic value. Second, we prove that the network's utilitarian value is constituted by its digital infrastructure characteristics: versatility, adaptability, interconnectedness and invisibility-in-use. The study is set within a software engineering company and bases on quantitative survey research, applying partial least squares structural equation modeling.

## Introduction

In conjunction with the popularity of public social media like Facebook or Twitter (Larosiliere, Carter, & Meske, 2017), the implementation of social media within organizational contexts has significantly increased over the last years and became an important focus of the information systems research community (Engler & Alpar, 2017; Soto-Acosta, Cegarra-Navarro, & Garcia-Perez, 2017). Enterprise social networks (ESNs) such as Confluence, IBM Connections, Jive, Slack, Sharepoint or Yammer have hence started to play an increasingly influential role at organizational workspaces. However, since participating in the provided ESN is usually not mandated, organizations often face the challenge of voluntary ESN adoption and continued use (Choudrie & Zamani, 2016; Engler & Alpar, 2017), why there is a high risk that ESN projects fail and lead to significant sunk costs. From literature on public social media we know, that hedonic motivations like enjoyment are recurrently mentioned as important factors for users to adopt or continually use social networks (e.g. Chen & Sharma, 2013; Hu, Poston, & Kettinger, 2011; Song, Wang, Zhang, & Qiao, 2017). At the same time, literature on information systems at the work place has shown, that usefulness is one of the most important predictors for use behavior (e.g. Davis, Bagozzi, & Warshaw, 1989; Lu, Yao, & Yu, 2005; Venkatesh & Davis, 2000; Venkatesh, Thong, & Xu, 2016). This relationship has also been proven in the context of ESN (e.g. Engler & Alpar, 2017). However, to the best of our knowledge, no comparison has yet been made between the role of utilitarian and hedonic motivations to use an ESN, which can be used for work purposes but also imports basic principles from public social media and therefore may be seen as a mean to socialize and increase enjoyment. Our first goal is to close this research gap by comparing the impact of perceived usefulness and perceived enjoyment on employees' ESN use continuance intentions.

Our second goal is to identify factors, that influence the employees' perception of the ESN's usefulness, as this aspect has two major implications for information systems managers. If they do not know why the ESN is useful, it is difficult to justify a corresponding project to the top management (Pérez-González, Trigueros-Preciado, & Popa, 2017; Wehner, Ritter, & Leist, 2017), what may impede introducing the ESN in the first place. In addition, managers face the challenge to communicate the work-related added value of the ESN to the potential participants. Shortcomings may lead to impeded ESN integration, or limited acceptance and usage (Iglesias-Pradas, Hernández-García, & Fernández-Cardador, 2017; Riemer, Stieglitz, & Meske, 2015). While ESNs have often been described as communication or collaboration infrastructures in work contexts (Kwahk & Park, 2016), an empirical investigation of infrastructure characteristics as antecedents of their usefulness is still missing. To fill this gap, we conceptualize ESNs as digital infrastructures and argue that certain infrastructure characteristics, which can also be found for physical infrastructures such as railways or roads, make ESNs useful for participants. This perspective helps in understanding that ESNs can be perceived as useful by employees, even though they are not designed to directly support certain business processes or specific tasks. In the current information systems literature, the term 'infrastructure' has been used to describe a multitude or diversity of hardware devices, software applications, or all-embracing application layers (Pipek & Wulf, 2009). According to Pipek and Wulf (2009) and based on the work of Star and Bowker (2002), information systems can be reconceptualized as digital work infrastructures, and can be described by certain key characteristics: interconnectedness, adaptability, reflexivity, versatility and invisibility-in-use. In this study, we examine how the perception of these infrastructure characteristics influences employees' perceived usefulness as an antecedent of ESN use continuance. Understanding the nature of ESNs is crucial, since they influence performance, particularly of knowledge-centric organizations (Cleveland & Ellis, 2014; Fischbach, Gloor, & Schoder, 2009; Leon, Rodríguez-Rodríguez, Gómez-Gasquet, & Mula, 2017; Wu & Chang, 2013).

In sum, we therefore pursue to answer the following two research questions:

(1) *To which extent do both factors, usefulness and enjoyment, influence the use continuance of enterprise social networks in organizations?*
(2) *How do the infrastructure characteristics of enterprise social networks influence employees' perception of the networks' usefulness and hence utilitarian value?*

To answer these two questions, in our study, we conduct a quantitative survey in an international, medium-sized company that offers software engineering and cloud services, and which is one of the leading companies in enterprise content management systems in Germany. In 2013, Confluence (developed by Atlassian) was introduced and promoted as a tool to support day-to-day work routines.

The contributions of this study are threefold. *First*, to the best of our knowledge, no study has directly compared the influence of perceived usefulness and enjoyment on ESN adoption or post-adoption behavior within an organization, even though both can play a significant role. In this regard, we fill a research gap regarding the phase of post-adoption, which has also important implications for practice. As we will show, a single-sided-analysis, e.g. utilitarian motivations only, would not take into account, that employees could also be motivated by hedonic aspects of ESN use at the same time, and vice versa, which eventually influences ESN implementation processes. *Second*, we adopt an infrastructure perspective on ESN that has often been discussed on a theoretical level but has not been tested empirically until now. Besides adding first empirical work in this context, our shift in perspective also impacts our general understanding of communication and collaboration technologies, such as ESNs, as rather being open work-infrastructures than task-oriented artefacts. This in turn influences research on the design, adoption and management of corresponding infrastructures, for which we however need a proper understanding of their core characteristics. This has also significant impacts on practitioners, since the shift of perspective on ESNs as digital infrastructures and better understanding of their perception by employees may influence corresponding investment, selection, management and evaluation processes within organizations. *Third*, we introduce new constructs to capture infrastructure characteristics, which can also be applied for adoption and post-adoption studies of other information systems, inside and outside of organizations.

The remainder of the paper is structured as follows: in the next section, we provide an overview of related work on ESNs in organizations, and describe both the utilitarian and hedonic aspects of ESN usage. We also introduce the perspective of ESNs as digital infrastructures. In the following section, the overall research model and hypotheses are developed. Afterwards, the research design is described in detail, including context information and measurement instruments. Then, we describe the results, including the descriptive statistics, measurement analysis and hypotheses testing. Subsequently, the findings are discussed and implications for research and practice are derived. This paper ends with a conclusion and outlook in terms of further research.

## Literature review

In this section, we discuss the role of ESNs in organizations. We also show the relevance of both the utilitarian and hedonic motivations for ESN use continuance. We then review the literature supporting the conceptualization of ESNs as digital infrastructures and how the infrastructural characteristics may influence the perceived usefulness of ESNs.

### Enterprise social networks in organizations

ESNs are defined as "platform[s] for tight integration of multiple types of Web 2.0 tools into a single private/semi-private network for businesses and organizations" (Scott, Sorokti, & Merrell, 2016, p. 2), and describe the phenomenon of social networking in an enterprise context via social media platforms. Typical examples that have been researched in context of information systems research, are for instance collaborative platforms such as Yammer, Jive, or Sharepoint (Choudrie & Zamani, 2016; Pawlowski et al., 2014; Riemer et al., 2015). ESNs can functionally be compared to social network sites (SNS) (e.g. Facebook, Twitter, LinkedIn) but differ in terms of accessibility, since these platforms can only be reached through the intranet of the enterprise (Richter, Riemer, & Vom Brocke, 2011). Thus, the majority of ESNs are only accessible exclusively by the employees of the respective organization (Richter et al., 2011). These organizational information systems are often implemented as stand-alone platforms or as tools which enhance organizational work processes through social network specific functionalities such as blogs, wikis and discussion forums (Turban, Bolloju, & Liang, 2011). Similarly to ordinary social network sites, ESNs commonly offer a core set of features which allow their users to e.g. create customized online profiles (Dugan et al., 2008); connect with other coworkers and track their activities (Wu, DiMicco, & Millen, 2010); share content and experiences by exchanging short messages with coworkers via direct messages or blogs (Richter et al., 2011; Riemer et al., 2015; Sarrel, 2010) or posting, commenting, editing and linking files with themselves or others (Leonardi, 2011). Additionally, these platforms offer enterprise specific collaboration capabilities (e.g. integration of existing communication tools, storage of documents or knowledge searches) which are not provided by common SNSs by default (Avanade, 2013). In the literature, ESNs are also referred to as enterprise social media (Beck, Pahlke, & Seebach, 2014), enterprise social networking systems (Fulk & Yuan, 2013), enterprise social software (Le-Nguyen, Guo, & Qiong, 2017), or organizational social web sites (Raeth, Kügler, & Smolnik, 2011).

ESNs are primarily used for internal communication purposes and to increase social interaction within the enterprise (Choudrie & Zamani, 2016). They are not typically linked to business processes or designed as purpose-driven artefacts such as ERP systems. Hence, there is an ongoing discussion on the utilitarian usefulness of ESNs. While current research in the field of information systems has already started to investigate the associated potential of ESNs in context of workspaces (Dimicco et al., 2008; Kwai Fun & Wagner, 2008; Leftheriotis & Giannakos, 2014; Moqbel & Aftab, 2015), the latest literature implies that an ESN's usefulness is influenced in a more indirect way, for example: i) quicker access to information and knowledge (Alpar, Engler, & Schulz, 2015; Engler, 2014); ii) increased group performance (Kügler, Smolnik, & Kane, 2015b; Wehner, Falk, & Leist, 2017); and iii) the establishment of intangible social capital (Cummings & Dennis, 2016; Riemer, Finke, & Hovorka, 2015). Particularly in terms of information sharing and knowledge transfer, organizational social media platforms have been identified as useful artefacts (Alsayadi & Algarni, 2017; Meske, Brockmann, Wilms, & Stieglitz, 2014; von Krogh, 2012). According to previous research investigations, the implementation of ESN technologies "can increase the accuracy of people's meta-knowledge (knowledge of 'who knows what' and 'who knows whom') at work" (Leonardi, 2015). As stated by Leon et al. (2017), the use of an internal social network tends to evaluate the knowledge flow within organizations and to determine which users act as knowledge diffusers by sharing what they know with others.

Beyond such knowledge-based investigations, ESNs seem to positively influence users' communication and group work (Raeth et al., 2011). Related work by Kwahk and Park (2016) demonstrates, for example, that ESNs can help to increase users' communication and collaboration. Users who are physically separated or do not share the same cultural profile can be connected via an ESN (Shirky, 2008). Besides these physical and cultural barriers, ESNs also have the potential to reduce hierarchical barriers within enterprises. For instance, communication clusters, in which users only communicate with other users of equal position within the enterprise hierarchy, seem to dissolve over time (Riemer et al., 2015). According to Beck et al. (2014) ESNs can be regarded as sociotechnical systems which provide a better sense of the social identity of others and increase the interaction transparency of users. Moreover, ESNs can potentially strengthen the relationships among users and enterprises (Fulk & Yuan, 2013).

In addition to enhanced group work experiences, social networking among employees supports creativity within groups, which may lead to competitive advantages (Yuhashi & Iijima, 2010). These findings are in line with recent investigations by Kügler et al. (2015b), who show that both the innovation and performance of employees are influenced by the way ESNs are used in practice. Hence, ESNs that are used for team building have a stronger effect on the task performance of employees. ESNs that are increasingly used for connecting teams have a stronger effect on employees' innovation. In line with these findings, research shows (Moqbel & Aftab, 2015; Moqbel, Nevo, & Kock, 2013) that the use of an employee social networking site leads to higher organizational commitment and job satisfaction. The authors also demonstrate the effect of ESN use on job performance through the mediation of job satisfaction (Moqbel et al., 2013).

### Enterprise social networks for utilitarian and hedonic purposes

Even though previous studies confirm the effect between the use of ESNs by employees and work performance (Kügler et al., 2015b; Leftheriotis & Giannakos, 2014; Moqbel et al., 2013), many companies are still facing issues such as limited employee usage. Although there has been research in the field of ESN acceptance in an organizational context, there are still unanswered questions concerning how to achieve sustainable employee participation (Engler & Alpar, 2017).

In traditional information systems adoption literature and the context of work performance, the measurement of perceived usefulness (Davis, 1989) or performance expectancy (Venkatesh, Morris, Davis, & Davis, 2003) has become a major determinant of adoption and post-adoption use continuance. Perceived usefulness is defined by Davis (1989, p. 320) as "the degree to which a person believes that using a particular system would enhance his or her job performance". According to prior literature, such utilitarian value is one of the strongest determinants of user acceptance in system-use environments (Lu et al., 2005; Venkatesh & Davis, 2000; Venkatesh et al., 2016). This seems also to apply to ESNs, as previous studies identified perceived usefulness or performance expectancy as significantly influencing the adoption of ESNs by employees (e.g. Antonius, Xu, & Gao, 2015; Engler & Alpar, 2017).

However, while highlighting the important accomplishments of these investigations, studies have also criticized an incomplete understanding of system-use behavior (e.g. Benbasat & Barki, 2007; Wu & Lu, 2013). One critical aspect of models and studies that focus on usefulness is that they may miss an important aspect in the context of systems that could also be used for hedonic purposes (Lin & Bhattacherjee, 2010; Wu & Lu, 2013). While perceived usefulness was shown to measure utilitarian value, it "may not be the sole prominent determinants for using hedonic systems" (Wu & Lu, 2013, p. 1). Utilitarian systems are designed to provide instrumental value, such as increasing the task performance or productivity of users. As defined by Sun and Zhang (2006) a system can be categorized as utilitarian when "it is aimed mainly at outcome-oriented tasks, in other words, when its users are mainly driven by an external locus of causality" (p. 622). In contrast, hedonic systems provide self-fulfilling value to the users (e.g. users experience enjoyment when using the system) (van der Heijden, 2004). If a system is categorized as hedonic, it "supports tasks focusing mainly on the process, and users have an internal locus of causality" (Sun & Zhang, 2006, p. 622). Despite their different natures, utilitarian and hedonic systems do not necessarily conflict with each other, and may be used for both productivity and pleasure (Chesney, 2006; Sun & Zhang, 2006; Wu & Lu, 2013).

From internet research on public social media we learn that social network use and continuous participation is strongly influenced by hedonic motivations (e.g. Chen & Sharma, 2013; Hu et al., 2011; Larosiliere, Meske, & Carter, 2015; Song et al., 2017). Transferring this finding to the workplace context within a company, recent qualitative investigations imply that the use of an ESN by employees may also be affected by hedonic motivations, not just by utilitarian values (e.g. Chin, Evans, & Choo, 2015). In addition, Kügler and Smolnik (2014) also showed that besides consumptive and contributive use, ESN may also be leveraged for hedonic and social use in the post adoption phase. Yet the impact (and comparison) of according motivations on future usage intentions has not been investigated. Further confirmation based on quantitative studies in the workplace context is still missing. Since individuals may use them either for utilitarian or hedonic purposes, ESNs could hence be categorized as dual-purpose systems as defined by Wu and Lu (2013). In our quantitative study, we therefore investigate at first to which extent usefulness as well as hedonic motivations influence ESN use behavior and compare the roles of both latent variables. In addition, while reasons for ESNs to satisfy hedonic needs (e.g. socializing) seem to be intuitive, there is still an ongoing discussion regarding the reasons for ESNs to be useful in the employees' day-to-day work routines. To shed light on the latter question, in the next part of this work, we conceptualize ESNs as digital infrastructures and argue that their infrastructure characteristics are the reason why users perceive ESNs as useful, even though they are not conventional, task-oriented information systems artefacts.

### Enterprise social networks as digital infrastructures

Recent investigations have demonstrated that besides the organizational, social and individual factors, technological factors also seem to have a strong impact on the utilitarian use of ESNs (Chin et al., 2015). These technological factors are "related to ESN platform characteristics" (Chin et al., 2015, p. 5) and to IT infrastructure, which is one of the key concerns in technology-related system contexts. IT infrastructure can serve as a strong driver, and can serve as an enabling or hindering factor in obtaining organizational competitive performance (Broadbent, Weill, Brien, & Neo, 1996). Weill and Vitale (2002) define IT infrastructure as "a set of services that users can understand, draw upon, and share, to conduct their business" (p. 19). As defined by Hwang, Yeh, Chen,

Jiang, and Klein (2002), IT infrastructure is "the base foundation for building business applications, which is shared throughout the firm as reliable services" (p. 56). While both definitions focus on a service perspective, this can be understood in the context of an ESN as a (new) work-oriented infrastructure for employees. According to the definition of Hanseth and Lundberg (2001), "work-oriented infrastructures are shared resources for a community; the different components of an infrastructure are integrated through standardized interfaces; they are open in the sense that there is no strict limit between what is included in the infrastructure and what is not, and who can use it and for which purpose or function; and they are heterogeneous, consisting of different kinds of components – human as well as technological." This technical perspective shows that infrastructures are not equal to standardized task-orientated artefacts. To properly design and manage the adoption process of a new work-orientated infrastructure, its basic characteristics need to be analyzed. According to Pipek and Wulf (2009), and similarly to Star and Bowker (2002), the following exemplary key characteristics need to be considered:

- *Versatility*: Infrastructures such as ESN can be used for many purposes. They are open systems that allow users to decide individually on utilitarian and hedonic usage scenarios that are unforeseen by management and even by the users themselves.
- *Invisibility-in-use*: When using an infrastructure, it remains "invisible" to the uers as long as everything works properly (front-end easy to use, back-end without disruptions); only if the infrastructure fails it becomes "visible", being cognitively present to the user.
- *Adaptability*: ESNs are based on a variety of different technology layers (e.g. organizational intranet, local area networks, the internet). The ability of the ESN to adapt to the IT environment is crucial to making the infrastructure work.
- *Reflexivity*: ESNs as digital infrastructures are often re-designed and modified on the basis of usage experiences. In this sense, the user becomes an essential part of the infrastructure design, thus ending the strict separation between users and designers. Furthermore, users can "modify and appropriate different parts of the [technology] in ways unforeseen by the technology designers" (Pipek & Wulf, 2009, p. 6).
- *Interconnectedness*: Like other types of infrastructure, an ESN is interconnected, for instance when looking at the hardware (e.g. storage, processors and servers) or software (e.g. applications and databases). At the same time, the ESN is not only interconnected itself, but its usage also helps people in the organization to interconnect. We hence see interconnectedness not exclusively from a technical perspective.

These key characteristics allow the quality of a work-oriented, digital infrastructure to be described, and have already been used in different contexts such as crisis management infrastructures (White, Plotnick, Kushma, Hiltz, & Turoff, 2009) or cloud computing infrastructures (Stieglitz, Meske, Vogl, & Rudolph, 2014) to measure and improve the impact of technology. For a better understanding of the following infrastructure-related hypotheses, we want to point out, that we understand an ESN as a digital infrastructure for its own, which however can be interconnected with other systems and hence be integrated in an overriding infrastructure. Using an analogy from the offline-world, the ESN could be compared to a city with its own infrastructure, which is interconnected via highways with the infrastructure of other cities nearby. The user can hence perceive the ESN as a digital (sub-)infrastructure but also evaluate its interconnectedness with other systems.

## Hypotheses development

This section presents the development of our research model. We first describe how both the perceived usefulness and perceived enjoyment influence an individual's ESN usage intention (Section 3.1). We then explain how infrastructure characteristics can influence the perceived usefulness of the ESN (Section 3.2). The research model is summarized in Figure 1.

### Utilitarian and hedonic motivation for using ESNs

An employee's decision to use ESN "is associated with the perception of ESN in enabling them to achieve a certain goal and fulfil their needs" (Chin et al., 2015, p. 5). As 'dual-purpose systems' (Wu & Lu, 2013), ESNs serve both utilitarian and hedonic goals. The usage of a utilitarian system primarily involves the motivation to improve job performance or increase the efficiency of work-related tasks (Appelbaum, Kogan, Vasarhelyi, & Yan, 2017; Davis, 1989). As a vast body of literature did demonstrate, perceived usefulness is strongly connected to the intention to continuously use an information system (Davis, Bagozzi, & Warshaw, 1992; Venkatesh, Thong, & Xu, 2012). Hence, we theorize that utilitarian system usage will also be positively correlated with an individual's intention to continuously use an ESN. Thus, we state:

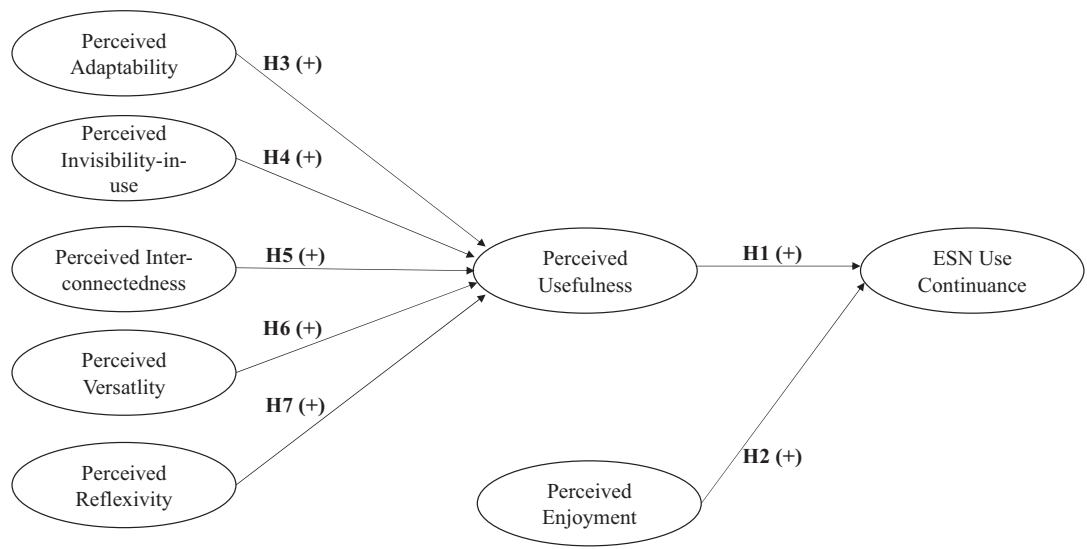

Figure 1. Research model.

*H1: Perceived usefulness is positively correlated with ESN use continuance*

ESNs are said to be affected by both hedonic and utilitarian values (Chin et al., 2015; Leftheriotis & Giannakos, 2014). Contrary to utilitarian systems, in hedonic systems the achievement of external goals is subordinated to the use of the system itself (van der Heijden, 2004). Perceived enjoyment is defined as the extent to which the activity of using computers is perceived to be enjoyable in its own right, apart from any performance consequences that may be anticipated (Alsabawy, Cater-Steel, & Soar, 2016; Davis et al., 1992; Moghavvemi, Sharabati, Paramanathan, & Rahin, 2017; Venkatesh & Davis, 2000). Hedonic motivations can be important both for the initial adoption of technology (Salehan, Kim, & Kim, 2017) and for intentions regarding continued use in post adoption models (Lowry, Gaskin, & Moody, 2015; Schwarz & Schwarz, 2014), as shown in the contexts of online shopping (Stafford & Stafford, 2001), streaming music (Chu & Lu, 2007) or videos (Kim, Na, & Ryu, 2007), experiencing virtual worlds (Wu, Li, & Rao, 2008), participating in public social networking or social blogging (Tscherning & Mathiassen, 2010). Thus, we theorize:

*H2: Perceived enjoyment is positively correlated with ESN use continuance*

### The influence of infrastructure characteristics on perceived usefulness

According to the layer approach and standardization (Pipek & Wulf, 2009), digital infrastructures are based on a variety of different protocols and can adapt to environments. These service layers are crucial to leveraging the IT infrastructure and ensuring it is adaptable to the existing environment. As demonstrated by Alsabawy et al. (2016), such infrastructure services are positively related to users' perceived usefulness. Hence, the perception of their adaptability may also influence the users' perceived usefulness of ESNs. We therefore hypothesize:

*H3: Perceived adaptability is positively correlated with the perceived usefulness of the ESN*

Another key characteristic of infrastructures is their invisibility-in-use. According to Star and Bowker (2002), infrastructures are invisible but become visible upon breakdown. This aspect can be explained with the example of roads or railways, which can be seen but are not 'consciously present' and hence 'visible' to the everyday work-traveler, until these infrastructures break down (he/she does not think about the roads/railways when commuting, the infrastructure is just 'there' in the background). Transferred to the information systems context, Pipek and Wulf (2009) argue that such invisibility-in-use also applies to information systems as infrastructures. We therefore adapt that understanding and assume, that using the ESN in e.g. daily communication-routines leads to the effect, that the visible artefact ESN becomes an 'invisible' but constantly available infrastructure, if it does not require high cognitive efforts to use it in daily work routines (front-end, interface) and if it runs smoothly without interruptions (back-end, server). Walker and Hong (2017) demonstrated that the presence of workplace infrastructure is related to users' perceived usefulness. Similar effects have already been shown in the context of e-learning technology, in which Wedel and Rothlauf (2014) demonstrated the mediated influence of infrastructure availability on perceived usefulness. As stated by Sun, Tsai, Finger, Chen, and Yeh (2008), "little operational reliability and long transmission times lead to learner frustrations and, hence, to negative emotions." Based on these findings, it is assumed that the perceived invisibility-in-use of an ESN is positively associated with users' perceived usefulness of the ESN.

*H4: Perceived invisibility-in-use is positively correlated with the perceived usefulness of the ESN*

The fifths hypothesis involves the interconnectedness of an infrastructure. Perceived interconnectedness describes the extent to which hardware and software layers are connected to other services to support an efficient work environment. One example that combines multiple infrastructures is grid computing; this leads to a reduction of administration effort since users no longer need to manage multiple standalone systems (Buyya & Sulistio, 2008; Strong, 2005). The interconnectedness of systems can also lead to a higher probability of IT adoption. ESNs are not only interconnected from a technical perspective but also allow users to interconnect with others, which can again increase their usefulness (Kügler, Dittes, Smolnik, & Richter, 2015a). The perceived interconnectedness, both technical and social, can play an important role in the users' perception of an ESN's usefulness. We hence hypothesize:

*H5: Perceived interconnectedness is positively correlated with the perceived usefulness of the ESN*

Another important criterion of infrastructure is versatility. According to Pipek and Wulf (2009), infrastructures can be used for various purposes in different work environments and areas, and can therefore be flexibly adapted to the respective business processes. The success of the infrastructure is measurable by its ability to combine several purposes in an instrument or technology (Morledge & Owen, 1997). The versatility of an infrastructure can hence have a positive influence on its usefulness. The literature has shown that ESNs can be used for many different purposes, such as to establish and access information repositories (Leon et al., 2017), contact experts (Han, Sörås, & Schjodt-Osmo, 2015) establish social capital (Beck et al., 2014), coordinate projects (Suh & Bock, 2015), socialize (Fulk & Yuan, 2013) and others. Hence, we assume that the perceived versatility of an ESN has a positive influence on its perceived usefulness.

*H6: Perceived versatility is positively correlated with the perceived usefulness of the ESN*

Infrastructures not only allow the user to store data (including data related to the infrastructure itself) and to identify applications, but are also subject to constant change, and are further developed and shaped on the basis of user experiences. The integration of end users' perceptions of the system development process therefore increases the perceived usefulness of a system (Foster & Franz, 1999). In this sense, the user becomes an essential part of the infrastructure design, and thus eliminates the strict separation between users and designers, resulting in the reflexivity of infrastructures. Information systems as reflexive infrastructures are part of the same global infrastructure as those of users and "all improvements to the global infrastructure are developed within that infrastructure" (Pipek & Wulf, 2009, p. 449). ESNs also undergo a reflexive development process through which the design and features are continuously enhanced (Wehner et al., 2017). We suggest that the perceived reflexivity of an ESN has a positive influence on the perceived usefulness of this work infrastructure. Hence, we hypothesize:

*H7: perceived reflexivity is positively correlated with the perceived usefulness of the ESN*

In the following figure, we summarize our research model.

## Research design

### Context information

The company of this study is a medium-sized organization with 260 employees. It was established in 1990, has its headquarter in Germany and operates internationally with branches in Austria, Great Britain, Turkey, the United States of America and Singapore. It is one of the leading companies in enterprise content management in Germany, and also offers software engineering and cloud services.

This company introduced Confluence (developed by Atlassian) as an ESN in 2013. Each employee has an account with a personal profile on the network. Whole departments are also represented on the network, e.g. finance and control, marketing and sales. The system offers classical ESN functionalities such as private messages, microblogs and wikis. Confluence was established in 2004 in Australia; it is one of the most popular social network solutions for small and medium-sized enterprises, and is available in 11 different languages. Since the management had a more conservative perspective and wanted to avoid, that the ESN is perceived as a duplicate of social media to only socialize and chit-chat, Confluence was introduced and promoted from the beginning as a tool to only support the employees' work tasks. Comparisons with public social media also obstructed the ESN investment decisions for some time because of the problem to calculate a return on investment for a tool, that somehow seems to support work-related tasks but is not designed as a task-driven artefact. In the following it is described, which measurement instruments were applied to investigate, if there is also a hedonic perspective besides the promoted utilitarian one that influences use continuance, and if an infrastructure perspective on ESN can help to increase the understanding of its perceived usefulness.

### Measurement instrument

In line with traditional information systems adoption methodology (Venkatesh et al., 2003, 2016), for our explanative approach we chose to conduct quantitative survey research within the boundaries of an organization and to apply structural equation modeling. The quantitative survey took place between October 4th and October 15th, 2016. It was advertised via e-mail to all employees, and participation in the study was optional. We adapted validated scales for use continuance from Agarwal and Karahanna (2000), which was used in many intra-organizational information systems contexts (e.g. Chandra, Srivastava, & Theng, 2012; Huang, Wu, & Chou, 2015) but also in the context of public social networks (e.g. Bataineh, Al-Abdallah, & Alkharabsheh, 2015). For perceived usefulness we adapted items from Davis et al. (1989), which is similar to the construct of performance expectancy (Venkatesh et al., 2003) and has also been applied in organizational ESN contexts (e.g. Engler & Alpar, 2017). For perceived enjoyment we adapted existing items from Davis et al. (1992), which has been applied in many studies ever since, also in the context of public social network adoption (e.g. Lin & Bhattacherjee, 2010). New (reflective) constructs were developed for the infrastructure characteristics of perceived versatility, adaptability, invisibility-in-use, interconnectedness and reflexivity. For this purpose, we followed the guidelines of Straub (1989). First, we derived initial item pools for each of the constructs and reduced the number of items successively by conducting multiple workshops (Cronbach, 1971) with up to four researchers. During these workshops, the participants were asked to evaluate the fit of the item for the target construct. We then conducted a quantitative pre-test with several Confluence users in the company, followed by qualitative interviews involving the individuals' understanding of the given items and constructs. Based on this feedback, the list of items was finalized. For each construct, a seven-point Likert scale was used, ranging from "strongly disagree" (1) to "strongly agree" (7). All measurement items are displayed in Table 1.

### Data analysis and results

### Descriptive statistics

In total, 121 of the 260 employees participated in the survey (response rate: 47%). Providing demographic data was optional. Of these, 102 of the participants were male (84%), 19 were female (16%). In terms of age, 37% (n = 45) were aged 40 or below, 31% (38) were between 41 and 50 years old, while 29% were between 51 to 60 years. One person was older than 60. A total of 88 people (73%) had been using the system for more than 12 months. A detailed overview of descriptive statistics is displayed in Table 2.

In line with information systems adoption theories (Bhattacherjee, 2001; Davis et al., 1989; Venkatesh et al., 2003, 2016), we assume that the experience with a system is a prerequisite for individual perceptions, e.g. regarding the system's usefulness, which are the cause for usage intentions and eventual post-adoption behavior. We tested, if experience moderates the correlation of perceptions of e.g. the ESNs usefulness with intention to use, but the results showed no significant moderating effect.

In the following Table 3, we provide an overview of the mean, min, max and standard deviation for all items of all constructs.

### Measurement model analysis

We applied partial least squares structural equation modeling (PLS-SEM) using SmartPLS 3.0. The constructs are reflective measures of their indicators. To assess convergent validity, we applied the following three criteria (Fornell & Larcker, 1981): the average variance extracted (AVE) for each construct should exceed .50 (see Table 4); the composite reliability (CR) of constructs should exceed .70 to assure construct reliability; and scale items should have loadings exceeding .70 on their respective scales. Our data met all criteria for convergent validity (see Table 4). Regarding internal consistency, all items factor loadings exceeded .70 (Bearden, Netemeyer, & Mobley, 1993) (see Table 4).

Regarding CR, all construct values exceeded .70 (see Table 4), and internal consistency can therefore be assumed (Bearden et al., 1993). In terms of discriminant validity, we compared the square root of the AVE of each construct, which was greater than any correlation in that construct's row or column, as recommended by (Fornell & Larcker, 1981) (see Table 4). In addition, the respective loadings were all lower than the cross-loadings (Gefen & Straub, 2005). Multicollinearity was tested using the variance inflations (VIF), resulting in values between 1.18 and 2.58, which are lower than the suggested maximum values of 5.00 (Menard, 1995).

**Table 1.** Definitions and measurement items.

| | |
|---|---|
| **Perceived Versatility (VER)**: The degree to which the user perceives the ESN as being useful for different purposes. (self-developed) | |
| VER1 | <The ESN> can be used for specific subtasks as well as superordinate tasks. |
| VER2 | <The ESN> can be used for many different work-related purposes. |
| VER3 | Different tasks can be performed in <the ESN>. |
| **Perceived Adaptability (ADA)**: The degree to which the user perceives the ESN as being adaptable to the given environment of the company. (self-developed) | |
| ADA1 | The structures in <the ESN> are constantly evolving. |
| ADA2 | <The ESN> structures are constantly adapting to the environment of the organization. |
| ADA3 | <The ESN> has become adapted to my organization's structures. |
| **Perceived Invisibility-in-use (INV)**: The degree to which the user perceives the ESN to be operating in the background without consciously noticing it. (self-developed) | |
| VIS1 | I often use <the ESN> unconsciously, as it has become an integral part of my working infrastructure. |
| VIS2 | I unconsciously use <the ESN> for many of my tasks. |
| VIS3 | <The ESN> is always available in the background during working hours. |
| **Perceived Interconnectedness (INT)**: The degree to which the ESN is interconnected with the existing processual, technological and social infrastructures. (self-developed) | |
| INT1 | In my organization, <the ESN> supports <1) only a few … 7) most> of the existing processes. |
| INT2 | In my organization, <the ESN> is an <1) important … 7) unimportant> component of the existing infrastructure. |
| INT3 | Through <the ESN>, I am connected with my colleagues. |
| **Perceived Reflexivity (REF)**: The degree to which the user perceives his/her activities as having an influence on the ESN's development over time. (self-developed) | |
| REF1 | <The ESN> allows me to influence its future development. |
| REF2 | I can participate in designing <the ESN>. |
| REF3 | My activities in <the ESN> have an influence on its evolving structures. |
| **Perceived Usefulness (PEU)**: The degree to which the individual evaluates the ESN as useful. (adapted from Davis, 1989) | |
| PEU1 | Using <the ESN> has been beneficial to me. |
| PEU2 | Compared to other resources, it is easier to access information through <the ESN>. |
| PEU3 | Using <the ESN> supports me in accomplishing tasks more quickly. |
| **Perceived Enjoyment (ENJ)**: The degree to which the user perceives the usage of the ESN as enjoyable. (adapted from Davis et al., 1992) | |
| ENJ1 | I find using <the ESN> to be enjoyable. |
| ENJ2 | The process of using <the ESN> is pleasant. |
| ENJ3 | I have fun using <the ESN>. |
| ENJ4 | Using <the ESN> gives me a good feeling. |
| **ESN Use Continuance (EUC)**: The intention of the user to continuously use the ESN. (adapted from Agarwal & Karahanna, 2000) | |
| EUC1 | I intend to continue actively using <the ESN> rather than discontinue its use. |
| EUC2 | My intentions are to continue using <the ESN> rather than to use any alternative means. |
| EUC3 | If I could, I would like to discontinue my use of <the ESN> (reverse coded). |

**Table 2.** Descriptive data.

| Gender | N | % |
|---|---|---|
| Male | 102 | 84.3 |
| Female | 19 | 15.7 |
| **Age** | | |
| <30 | 16 | 13.2 |
| 31–40 | 29 | 24.0 |
| 41–49 | 38 | 31.4 |
| 51–60 | 35 | 28.9 |
| >60 | 1 | .80 |
| No response | 2 | .70 |
| **ESN Usage Experience (months)** | | |
| 0–12 | 17 | 14.0 |
| 13–18 | 26 | 21.5 |
| 19–24 | 17 | 14.0 |
| >24 | 45 | 37.2 |
| No response | 16 | 13.2 |

**Table 3.** Descriptive statistics of item values.

| | Mean | Min | Max | Std. Dev. | | Mean | Min | Max | Std. Dev. |
|---|---|---|---|---|---|---|---|---|---|
| **Perceived Versatility (VER)** | | | | | **Perceived Reflexivity (REF)** | | | | |
| VER1 | 4.91 | 2 | 7 | 1.34 | REF1 | 3.06 | 1 | 7 | 1.43 |
| VER2 | 5.30 | 1 | 7 | 1.29 | REF2 | 2.63 | 1 | 7 | 1.40 |
| VER3 | 4.93 | 1 | 7 | 1.49 | REF3 | 2.95 | 1 | 7 | 1.49 |
| **Perceived Adaptability (ADA)** | | | | | **Perceived Usefulness (PEU)** | | | | |
| ADA1 | 5.06 | 2 | 7 | 1.29 | PEU1 | 3.93 | 1 | 7 | 1.70 |
| ADA2 | 4.86 | 2 | 7 | 1.40 | PEU2 | 4.96 | 1 | 7 | 1.55 |
| ADA3 | 4.15 | 1 | 7 | 1.51 | PEU3 | 4.06 | 1 | 7 | 1.74 |
| **Perceived Invisibility-in-use (INV)** | | | | | **Perceived Enjoyment (ENJ)** | | | | |
| VIS1 | 3.97 | 1 | 7 | 1.93 | ENJ1 | 3.88 | 1 | 7 | 1.65 |
| VIS2 | 3.63 | 1 | 7 | 1.70 | ENJ2 | 4.06 | 1 | 7 | 1.54 |
| VIS3 | 4.92 | 1 | 7 | 1.79 | ENJ3 | 3.75 | 1 | 7 | 1.64 |
| | | | | | ENJ4 | 3.04 | 1 | 7 | 1.50 |
| **Perceived Interconnectedness (INT)** | | | | | **ESN Use Continuance (EUC)** | | | | |
| INT1 | 4.30 | 1 | 7 | 1.41 | EUC1 | 5.27 | 1 | 7 | 1.52 |
| INT2 | 4.98 | 1 | 7 | 1.19 | EUC2 | 4.83 | 1 | 7 | 1.78 |
| INT3 | 5.65 | 1 | 7 | 1.34 | EUC3 | 5.09 | 1 | 7 | 1.95 |

## Hypotheses testing

We estimated the structural model using PLS (Ringle, Wende, & Becker, 2015). R2 for ESN use continuance is 65%, suggesting substantive data variation that is explained by the independent variables of perceived usefulness and perceived enjoyment. Perceived usefulness has a positive and significant effect (β = .54; $p < .001$) on ESN use continuance, which means that Hypothesis 1 is supported. Perceived enjoyment also

**Table 4.** Measurement model analysis and inter-construct correlations.

|     | CR  | AVE | ADA | VER | INV | REF | INT | PEU | ENJ | EUC |
| --- | --- | --- | --- | --- | --- | --- | --- | --- | --- | --- |
| ADA | .88 | .70 | **.84** |     |     |     |     |     |     |     |
| VER | .92 | .79 | .62 | **.89** |     |     |     |     |     |     |
| INV | .88 | .70 | .53 | .55 | **.84** |     |     |     |     |     |
| REF | .87 | .69 | .39 | .32 | .36 | **.83** |     |     |     |     |
| INT | .82 | .69 | .37 | .28 | .29 | .18 | **.83** |     |     |     |
| PEU | .93 | .81 | .62 | .58 | .67 | .42 | .45 | **.90** |     |     |
| ENJ | .96 | .85 | .60 | .78 | .62 | .47 | .35 | .78 | **.92** |     |
| EUC | .89 | .73 | .55 | .56 | .62 | .35 | .34 | .78 | .73 | **.86** |

CR = composite reliability; AVE = average variance extracted; ADA = adaptability; VER = versatility; INV = invisibility-in-use; REF = reflexivity; INT = interconnectedness; PEU = perceived usefulness; ENJ = perceived enjoyment; EUC = ESN use continuance

**Table 5.** Direct and indirect effects.

| Predictor | Outcome | Standardized β | Standard Error | t-value | p-value |
| --- | --- | --- | --- | --- | --- |
| *Direct effects* | | | | | |
| ADA | PEU | 0.21 | 0.08 | 2.68 | .004 |
| VER | PEU | 0.14 | 0.07 | 2.06 | .020 |
| INV | PEU | 0.39 | 0.07 | 5.70 | .000 |
| REF | PEU | 0.11 | 0.08 | 1.39 | .082 |
| INT | PEU | 0.20 | 0.12 | 1.68 | .047 |
| PEU | EUC | 0.54 | 0.09 | 6.19 | .000 |
| ENJ | EUC | 0.31 | 0.10 | 3.21 | .001 |
| *Indirect effects* | | | | | |
| ADA | EUC | 0.12 | 0.04 | 2.69 | .004 |
| VER | EUC | 0.08 | 0.04 | 1.89 | .029 |
| INV | EUC | 0.21 | 0.05 | 4.02 | .000 |
| REF | EUC | 0.06 | 0.04 | 1.40 | .080 |
| INT | EUC | 0.11 | 0.07 | 1.50 | .067 |

CR = composite reliability; AVE = average variance extracted; ADA = adaptability; VER = versatility; INV = invisibility-in-use; REF = reflexivity; INT = interconnectedness; PEU = perceived usefulness; ENJ = perceived enjoyment; EUC = ESN use continuance has a significant and positive effect (β = .31; $p < .01$) on ESN use continuance, supporting Hypothesis 2.

The $R^2$ of perceived usefulness is 61%, suggesting high data variation that is explained by the independent constructs of infrastructure characteristics. Perceived adaptability has a significant and positive effect (β = .21; $p < .01$) on perceived usefulness, supporting Hypothesis 3. Perceived invisibility-in-use (β = .39; $p < .001$) also shows a highly significant and positive path to perceived usefulness, supporting Hypothesis 4. Perceived interconnectedness (β = .20; $p < .05$) has a significant and positive effect on perceived usefulness, as does versatility (β = .14; $p < .05$), supporting Hypothesis 5 and 6. Perceived reflexivity has a positive but insignificant correlation (β = .11; $p > .05$) with perceived usefulness, and hence Hypothesis 7 was not supported. All direct effects are stronger than the indirect effects (see Table 5).

A summary of our estimated final model is illustrated in Figure 2, the results regarding the support of the hypotheses are summarized in Table 6.

To assess the statistical power of our data, we analyzed the effect size by calculating Cohen's $f^2$. Cohen (1988) suggested the following criteria for interpreting effect size: (i) for a small effect, $.02 < f^2 \leq .15$; (ii) for a medium effect, $.15 < f^2 \leq .35$; and (iii) for a large effect, $f^2 > .35$. The size of the effect of perceived usefulness was large ($f^2 = .33$) and contributed significantly to the $R^2$ of use continuance. The effect size for perceived enjoyment is .10 and hence small. Regarding the contribution to $R^2$ of perceived usefulness, the effect size of perceived invisibility-in-use is large ($f^2 = .23$). Regarding the contribution to the $R^2$ of perceived usefulness, low effect sizes were found for perceived adaptability ($f^2 = .06$), perceived versatility ($f^2 = .03$), perceived reflexivity ($f^2 = .03$) and perceived interconnectedness ($f^2 = .08$). In addition, we analyzed the predictive relevance of our model by application of the Stone-Geisser test ($Q^2$), which indicates how well the data can be reproduced by the PLS model. The Q2 values for individual performance impact (Q2 = .45) and ESN use continuance (Q2 = .44) are positive, indicating a high level of predictive relevance.

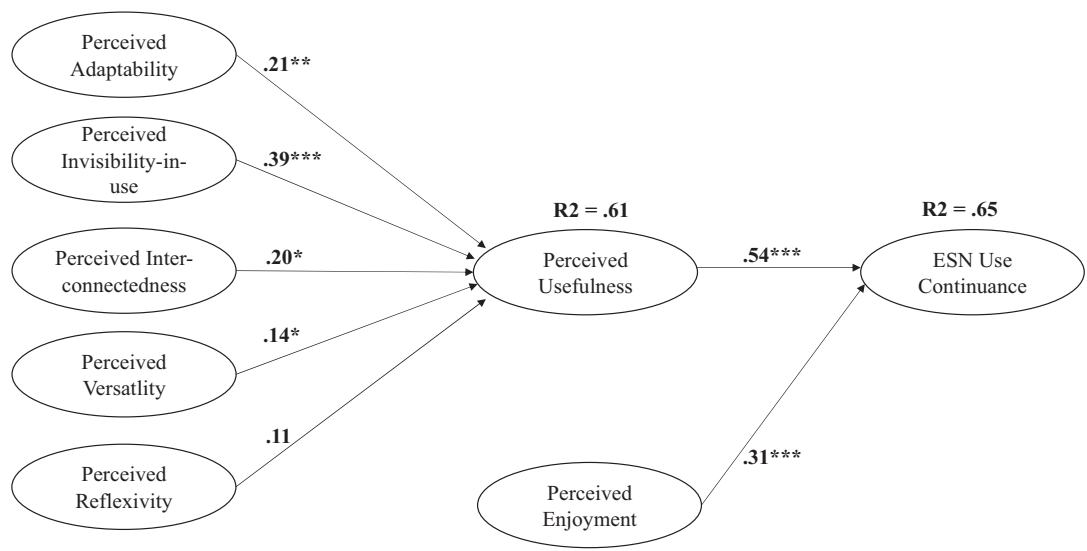

**Figure 2.** Model results.

Table 6. Hypotheses results.

| Hypotheses | Result |
| --- | --- |
| H1: Perceived usefulness is positively correlated with ESN use continuance | Supported |
| H2: Perceived enjoyment is positively correlated with ESN use continuance | Supported |
| H3: Perceived adaptability is positively correlated with the perceived usefulness of the ESN | Supported |
| H4: Perceived invisibility-in-use is positively correlated with the perceived usefulness of the ESN | Supported |
| H5: Perceived interconnectedness is positively correlated with the perceived usefulness of the ESN | Supported |
| H6: Perceived versatility is positively correlated with the perceived usefulness of the ESN | Supported |
| H7: Perceived reflexivity is positively correlated with the perceived usefulness of the ESN | Not Supported |

## Discussion and implications

Our analysis of ESNs as digital infrastructures yielded several results. In the following, we will discuss the results and thereby answer the two research questions, which were stated in the introduction.

### Utilitarian value outpaces hedonic motivations for ESN use continuance

The results show, that in the investigated company, both, usefulness as well as enjoyment have an important influence on the employees' intention to continue using the ESN, answering the first research question. First, the finding is in line with recent quantitative studies, which found that the utilitarian value is important for ESN adoption and use behavior, as shown by e.g. Engler and Alpar (2017). It is also in line with Kügler and Smolnik (2014) who showed in a survey-based study on ESN post-adoption behavior that the means for e.g. work-related contributive usage were higher than those for hedonic usage. Yet, our findings also indicate, that social networks, even if introduced by the management for utilitarian purposes only, seem to have an inherent hedonic aspect, which needs to be taken into account when implementing and managing the ESN. On a quantitative basis, we therefore prove the assumptions of the qualitative study by Chin et al. (2015) who theorized, that it is not only the utilitarian but also hedonic motivation, that influences ESN use continuance intentions. This is also in accordance with recent discussions on technology affordances, which state that information systems and hence ESN use may vary based on individual perception as well as contextual needs (Leidner, Gonzales, & Koch, 2018; Leonardi & Vaast, 2017). According to Affordance Theory, affordances are possibilities for action, which emerge as individuals interact with technologies (Hutchby, 2001; Stoffregen, 2003). These possibilities, may they be related to utilitarian or hedonic aspects, can differ within collective groups of individuals using the same communication and collaboration technologies, for instance within organizations. Depending on the particular workplace context in which workers have to communicate or exchange information with others, they may follow also hedonic motivations in utilizing ESN rather than act only to utilitarian goals of the management. Our results therefore demonstrate that with the introduction of ESNs, basic principles of public social media, which according to literature are mainly used for hedonic purposes (e.g. Chen & Sharma, 2013; Hu et al., 2011; Song et al., 2017), were also imported to the workplace area. It is hence necessary to acknowledge, that ESNs remain "social technologies" to a certain extent, even if specifically introduced and promoted to increase work-related access to information and knowledge only.

However, the results also show, that utilitarian motivations take on the majority of explanatory power for sustainable engagement in ESN. When we compare the standardized β, the path coefficient between perceived usefulness and use continuance is almost twice as high as the one of perceived enjoyment and use continuance. In addition, when comparing Cohen's $f2$ between perceived enjoyment and perceived usefulness, the statistical power of the latter is over three times as strong. First, these findings are in line with general information systems adoption literature, in which the utilitarian value of information systems is postulated to be the most important predictor for behavioral intentions and eventual use behavior (e.g. Bhattacherjee, 2001; Davis, 1989; Venkatesh et al., 2003, 2016). Second, if additionally taking the comparably high means of corresponding usefulness-items into account, it shows that such social technologies can be perceived as having an important utilitarian value for the employees' in day-to-day work routines. While these findings provide further indications that help to answer the often-discussed question *if* ESNs are useful, it yet does not provide information, *wh*y they are useful. In the next section, we therefore discuss the role of infrastructure characteristics for the perceived utilitarian value of ESNs in organizations.

### Enterprise social networks can be understood as digital infrastructures

Our analysis supports the assumption that infrastructural determinants influence the perceived usefulness of an ESN. Based on the integrated perspective theory on design and use of IT by Pipek and Wulf (2009), we can show that in our study, four infrastructural characteristics significantly influenced the employees' perception of the ESN's utilitarian value.

The results show that **perceived invisibility-in-use** has the strongest effect of all infrastructural criterions on perceived usefulness in terms of both, the path coefficient as well as statistical power. As an adequate digital infrastructure, the system needs to work fluently without interruptions, and, as a consequence, recedes into the background (Star & Ruhleder, 1996). As soon as errors occur, and the employees are no longer able to carry out their work properly, the ESN as an infrastructure becomes visible. It is difficult for designers to conceptualize and to customize the ESN infrastructure to increase the effect of invisibility-in-use. Bug fixes and improving design adjustments play an important role in the ongoing design process (Pipek & Wulf, 2009). Invisibility-in-use could also be compared to the concept of "immersion", in which cognitive absorption leads to users "diving into" the virtual collaboration environment (Agarwal & Karahanna, 2000; Vishal, 2016). It is therefore necessary, to increase immersive aspects of ESN use through its design and trouble-free operation.

**Perceived adaptability** is found to be another infrastructural criterion with a significant influence on perceived usefulness. This factor characterizes an infrastructure in terms of the degree to which the user perceives the system as being adaptable to the given environment of the company. In general, many different aspects in the organizational and technical environment constantly change in enterprises over time. If an ESN is perceived as being adaptable to according modifications, it helps to keep or even increase its usefulness. This development encompasses both technical and social elements, which are defined by Henfridsson and Bygstad (2013) as adaptation, innovation and scaling. In this context, adaptation means that the utilitarian value of the ESN as an infrastructure increases if e.g. the number of users increases, who can be incorporated by it. Innovation is defined as a self-reinforcing process through which new products and services extend the infrastructure, which in the case of an ESN would refer to its features and interfaces. Finally, the range of scaling increases due to the possibility of better collaboration via the system (Henfridsson & Bygstad, 2013).

**Perceived interconnectedness** is the third determinant that is found to significantly influence the perceived usefulness of the ESN. From a technical perspective, it is important that the users perceive the ESN as being interwoven with other necessary systems at work. This aspect can relate to interconnectedness in terms of both hardware and software in the organization. At the same time, from a social perspective, the ESN helps the users to be interconnected with their colleagues. A network can be established that can help to disseminate and access information and expertise quickly, without the barriers of departmental structures or hierarchies (Riemer et al., 2015). Interconnectedness is hence an important characteristic that impacts ESN use through increased perceived usefulness. In consequence, information systems managers need to consider possibilities to integrate the ESN into the existing infrastructural environment, linking it to other information systems. In addition, it is important to increase the perception of interconnectedness between the users, which has been proven to positively influence their individual work performance (Kügler et al., 2015a)

In terms of **perceived versatility**, our work shows a significant positive correlation with the perceived usefulness of the ESN. Pipek and Wulf (2009) state in their work that infrastructures can be used for different purposes. ESNs can also be applied in different work scenarios, with or without colleagues being involved, and can help to support tasks that were unforeseen by the ESN designers. To profit from this inherent versatility, employees need to be able to use the ESN in a self-determined and self-organized way, according to their own needs, responsibilities and work experiences. Only then can the ESN become versatile 'equipment' (Riemer & Johnston, 2017) that increases work efficiency and effectiveness, and hence positively influences its usage.

**Perceived reflexivity** is not found to have a significant impact on the utilitarian value of the ESN for the user. We were hence not able to prove that a user who believes that their activities in the ESN have an impact on the system's design and functionality perceives a higher utilitarian value for the infrastructure. The results are difficult to interpret, since there is yet no related work in this field. The current literature focuses mainly on the interaction between technology designers and users, and shows how the relationship between these actors can influence the design or the acceptance and use of technologies (Harris & Weistroffer, 2009; Park & Park, 2014).

In sum, *ESNs can be understood as digital infrastructures* that help employees to carry out their work more efficiently and effectively. This in turn affects the question of a potential return on investment. Due to its versatility, unforeseeable usage by employees and the fact that the ESN is usually not directly linked to business processes, it remains difficult to calculate a monetary added value for the ESN. However, the question of a return may not be a useful one; for example, one would not attempt to estimate the added value of a communication infrastructure such as a telephone system. Managers may have to relinquish the idea that ESN projects are comparable to conventional IT projects, in which the system usage can be controlled and a return can be quantified or even monetized. At the same time, work in knowledge-centric organizations has become more autonomous and anonymous, and in these environments ESNs have become crucial for self-organized workers to create and access informational paths that are unforeseeable by the managers. In addition, they help to keep employees in touch and increase the potential effects of serendipity. Management should therefore not promote an ESN as a specific tool for very specific tasks, but could provide exemplary ESN use cases and leave room for the employees to build their own usage scenarios and networks.

### Limitations

As in every research study, this work involves limitations due to its design choices and the nature of the dataset. We derive our findings from a single company, impacting the study's generalizability. Yet, the software engineering company can be seen as representing a knowledge-centric service industry that significantly relies on employee contributions. Moreover, the construct of interconnectedness reflects aspects that are not solely related to technical aspects, since it also captures the interconnectedness of employees. Even though this approach is based on multiple workshops to establish the construct, one may argue that zooming in on the technical aspects of interconnectedness could be more conclusive. We would hence advice to adjust and test the corresponding construct items. In addition, we focused on antecedents of perceived usefulness, not of perceived enjoyment. As a consequence of this study design, we did not test, if single items such as INT3 also correlate with perceived enjoyment. Also, the investigation was conducted in Germany, and the findings may not be directly transferable to cultures of other countries in non-Western areas. In addition, behavioral intention does not necessarily lead to actual usage. Unfortunately, we were not able to analyze if the individual's behavioral intention of continued use did lead to actual use behavior, because of impeding data privacy policies.

### Conclusion

In this work, we found that both, the utilitarian value as well as hedonic motivations influence ESN use continuance in organizations. We furthermore showed, that ESNs can be understood as digital infrastructures and that especially the attributes of adaptability, invisibility-in-use, interconnectedness and versatility of the ESN have a significant positive influence on its perceived usefulness. The results offer a new perspective on organizational technology, which can help to explain the gap between the perception of an ESN as being useful and the concurrent problem of measuring its added value. ESNs as digital infrastructures are flexible, constantly evolving enablers for multiple purposes, which cannot be defined beforehand. An added value for the individual and the company is an option rather than a promise.

With this study, we hence significantly increase the still limited understanding of the ESNs' "essence" by reconceptualizing them as digital infrastructures. This allows us to address questions that have been unanswered before, having a substantial impact on current and future ESN research as well as ESN management. Furthermore, we established and introduced a new instrument to analyze the perception of digital infrastructure characteristics, which did not exist before and may support future studies in any field of infrastructure research. In addition, we are also the first to directly compare the importance of usefulness and enjoyment of social technologies at the workplace for use continuance. Therefore, this study contributes to the information systems research and adjacent fields in general, but it also covers specific topics of the ESN community.

Future studies could benefit from a longitudinal approach that compares the behavioral intention with the actual, continuous engagement in the ESN. Also, future research should investigate the perception of ESNs as infrastructures in different companies of the same industry, between different industries as well as between different cultural settings. Also, the perceived infrastructure characteristics of different ESNs such as Confluence, IBM Connections, Jive, Slack, Sharepoint, Yammer or others could be analyzed. In addition to our quantitative investigation, future studies might add content analysis or qualitative interviews with both users and managers to learn more about the consequences that an infrastructure perspective on ESNs creates. In addition, a new approach on how to measure the value or return on investment of digital infrastructures would be of interest to researchers and practitioners.

### Notes on contributors

*Christian Meske* is Assistant Professor at the Department of Information Systems, Freie Universität Berlin and Einstein Center Digital Future, Germany. He received his PhD from the Department of Information Systems, University of Münster, Germany. His research on digital workplace technologies and transformation has been published in journals including *Business & Information Systems Engineering, Business Process Management Journal, Information Systems Frontiers, Journal of the Association for Information Science and Technology*, and *Journal of Enterprise Information Management*. Amongst others, Christian has been recognized with the AIS Best Information Systems Publication of the Year Award.

*Konstantin Wilms* is a research associate at the Department of Computer Science and Applied Cognitive Science, University of Duisburg-Essen (Germany). He studied Applied Cognitive and Media Science at the University of Duisburg-Essen (Germany) and is currently a PhD student at the research group for Professional Communication in Electronic Media/Social Media at University of Duisburg-Essen, Germany. His work on switching behaviour and IT discontinuance was published in various reputable IS conferences, including the International Conference on Information Systems.

*Stefan Stieglitz* is Professor at the Department of Computer Science and Applied Cognitive Science at University of Duisburg-Essen, Germany. His research interest is on socio-technical systems and user behavior in information systems. His work has been published in journals including *European Journal of Information Systems, Journal of Management Information Systems* and *International Journal of Information Management*. Amongst others, Stefan has been recognized with the AIS Best Information Systems Publication of the Year Award.

## ORCID

Christian Meske  http://orcid.org/0000-0001-5637-9433